\documentclass[osajnl,twocolumn,showpacs,superscriptaddress,10pt]{revtex4-1}

\usepackage{amsmath,amssymb,graphicx}

\begin{document}

\title{Off-axis QEPAS using a pulsed nanosecond Mid-Infrared Optical Parametric Oscillator}

\author{Mikael Lassen}\email{Corresponding author: ml@dfm.dk}
\affiliation{Danish Fundamental Metrology, Matematiktorvet 307, DK-2800 Kgs. Lyngby, Denmark}

\author{Laurent Lamard}
\affiliation{Laserspec BVBA, 15 rue Trieux Scieurs, B-5020 Malonne, Belgium}

\author{Yuyang Feng}
\affiliation{COPAC ApS, Diplomvej 381, Scion science park, DK-2800, Kgs. Lyngby, Denmark}

\author{Andre peremans}
\affiliation{Laserspec BVBA, 15 rue Trieux Scieurs, B-5020 Malonne, Belgium}

\author{Jan C. Petersen}
\affiliation{Danish Fundamental Metrology, Matematiktorvet 307, DK-2800 Kgs. Lyngby, Denmark}



\begin{abstract}
A trace gas sensor, based on quartz-enhanced photoacoustic spectroscopy (QEPAS), consisting of two acoustically coupled micro-resonators (mR) with an off-axis 20 kHz quartz tuning fork (QTF) is demonstrated. The complete acoustically coupled mR system is optimized based on finite element simulations and experimentally verified. The QEPAS sensor is pumped resonantly by a nanosecond pulsed single-mode mid-infrared optical parametric oscillator (MIR OPO). The sensor is used for spectroscopic measurements on methane in the 3.1 $\mu$m to 3.5 $\mu$m wavelength region with a resolution bandwidth of 1 cm$^{-1}$ and a detection limit of 0.8 ppm. An Allan deviation analysis shows that the detection limit at optimum integration time for the QEPAS sensor is 32 ppbv@190s and that the background noise is solely due to the thermal noise of the QTF.
\end{abstract}

\maketitle

\section{Introduction}
Photoacoustic spectroscopy (PAS) is a very promising method for environmental, industrial, and biological monitoring, due to its ease of use, compactness and its capability of allowing trace gas measurements at the sub-parts per billion (ppb) level \cite{Harren2000,Sigrist2003,Hodgkinson2013,Besson2006,Lassen2015,Peltola2015,Szabo2013}. The photoacoustic (PA) technique is based on the detection of sound waves that are generated due to absorption of modulated optical radiation. Quartz tuning forks (QTFs) have shown great potential as sound transducers for PAS and have been increasingly applied to selective and sensitive detection of trace gases since its introduction in 2002 \cite{Kosterev2002,Patimisco2014,Spagnolo2012,Jahjah2014,Kohring2015}. Standard low cost QTFs with resonance frequencies at 32.7 kHz are typically used as sensors, however, also custom made QTFs have been reported \cite{Patimisco2014,Spagnolo2015,Zheng2016}. The PA signal is proportional to the Q-factor: $S \propto Q P/f_0$, where $P$ is the optical power, $\alpha$ is the molecular absorption coefficient and $f_0$ is the resonant frequency of the QTF.  The stiffness of quartz provides an efficient means of confining the acoustic energy in the prongs of the QTF, resulting in large quality factors (Q-factors) of the order 100000 in vacuum and 3-8000 at atmospherical pressures \cite{Dong2010}. This enables therefore  detection of very weak PA excitation with very small gas volumes. The QTF is highly immune to environmental noise since the width of the resonance at normal pressure is 4-5 Hz, allowing only noise frequency components in this narrow spectral band to excite QTF vibrations. Typically QEPAS-based systems consist of a QTF coupled to a micro-resonator (mR) in order to enhance the PAS signal. The mR is formed by one or two thin tubes, and the QTF is positioned between or beside the mR tubes to probe the acoustic signal excited in the gas \cite{Liu2009,Ma2013,Yi2012,Yi2014,Dong2014,Zheng2016}. This kind of positioning is called on-axis and off-axis coupled, respectively. For an optimal QEPAS sensor the resonant frequencies of the mR and the QTF must be matched. Such a frequency-matching design is not straightforward, because the gas-filled mR tubes and the QTF are acoustically coupled and affect the resonant properties of each other \cite{Yi2012,Dong2010}.

In this letter we report on a novel QEPAS sensor design with two mR tubes, an on-axis mR tube acting as absorption cell and an off-axis mR tube acoustically coupled, where a 20 kHz QTF is placed. The off-axis mR system has many advantages compared with traditional on-axis QEPAS systems. It makes optical alignment easy and thus background free measurements are easier to achieve. We find that the resulting background noise signal is solely due to the QTF thermal noise. Another advantage is that it allows mechanical protection of the QTF, since only the top of the QTF casing needs to be opened. The QEPAS experiment is conducted with a nanosecond pulsed single-mode MIR OPO. The resonance of the QTF is excited resonantly by setting the repetition rate of the OPO to 19.99 kHz. Spectroscopic measurements are conducted on methane in the 3.1-3.5 $\mu$m wavelength region, where background free measurements and a sensitivity of 0.8 ppm are obtained.

\section{Simulation of the acoustic response}

\begin{figure}[htbp]
\centering
\fbox{\includegraphics[width=\linewidth]{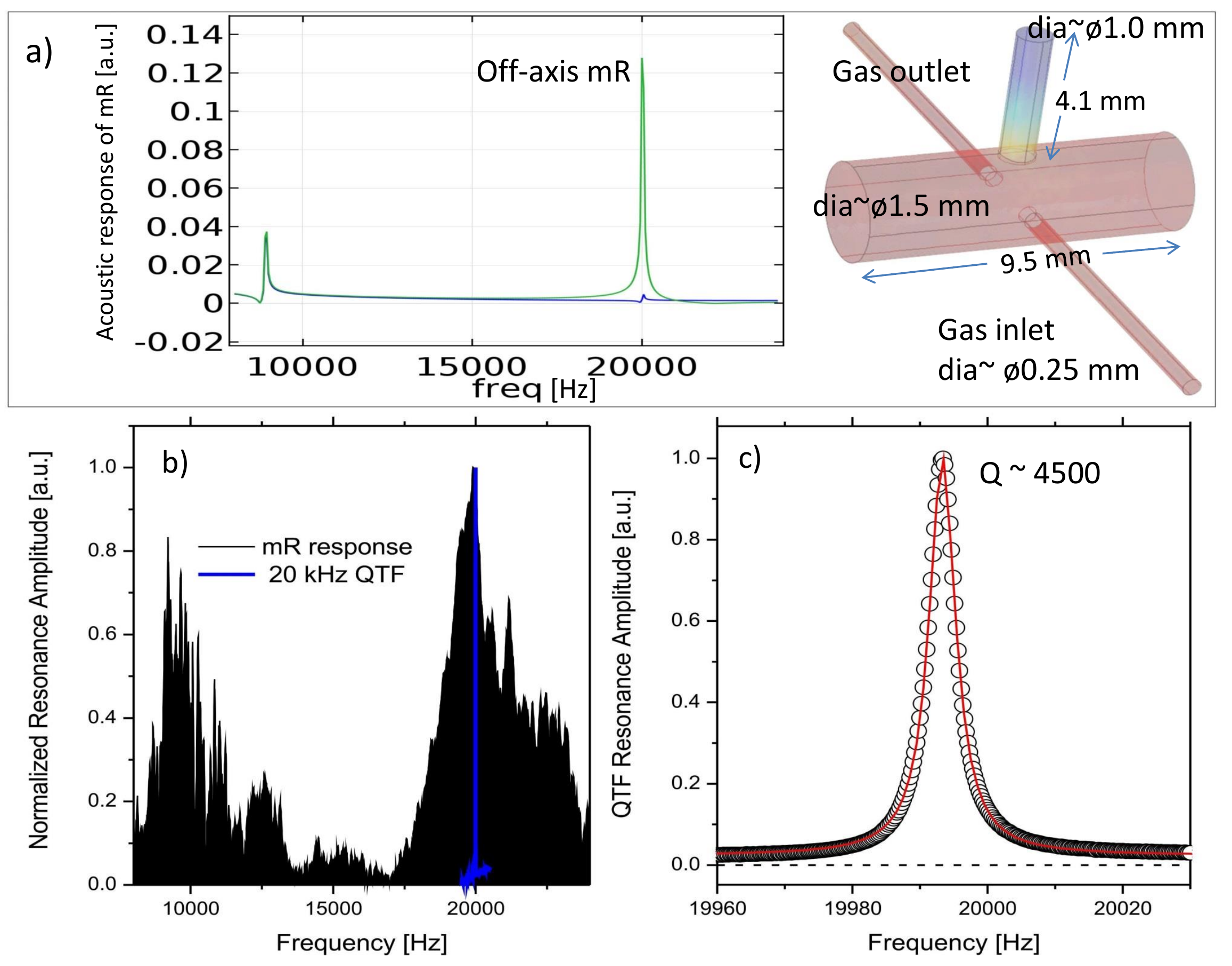}}
\caption{a) The left figure shows the simulations of the frequency response of the coupled acoustic mR system. The right figure shows the dimensions and design of the modeled QEPAS system. b) Experimental measured acoustic response of the coupled mR system (black curve) and the QTF (blue curve). The Q-factor of the 20 kHz resonance is approximately 8. c) Normalized frequency response of the QTF. Measured Q-factor $\sim$ 4500.}
\label{fig2}
\end{figure}

The acoustics response of the PA cell is simulated using finite-element simulations. Finite element simulations are a powerful tool for designing PA cells \cite{Lassen2014} and similar systems has already been successfully simulated \cite{Baumann2007}. The simulations allow the determination of the optimal positions of the QTF in the off-axis tube and the optimal dimensions of the coupled mR tubes in order to have the highest acoustic coupling to the QTF. The simulated frequency response of the coupled acoustic mR system is shown in Figure \ref{fig2}a). We find that the coupled system has two main eigenfrequencies one at 9 kHz and one at 20 kHz. The 9 kHz resonance has then same magnitude in both the on-axis tube and the off-axis tube. From the simulation we find that it is due to the gas inlet/outlet tubes and the coupling to the mR tubes, while the 20 kHz resonance is associated with the resonance of the off-axis tube as expected. In Figure \ref{fig2}a) the dimensions and design of the optimal design is shown. The on-axis mR (absorption tube) is 9.5 mm-long with 1.5 mm inner diameter coupled to the off-axis mR with a 4.1 mm long tube with 1 mm inner diameter, where the QTF is placed. The gas inlet/outlet tubes have a 0.5 mm inner diameter. This design seems to yield the best acoustic coupling to the QTF at 20 kHz. In order to, verify the simulation measurements are conducted with a microphone with a bandwidth of 20 kHz and a loudspeaker. The microphone was placed in the off-axis tube. Figure \ref{fig2}b) shows the measured acoustic resonance of the the coupled mR system.  The data has been normalized with the response transfer function of the microphone measured in free space. It can be seen that the system has two resonances one at approximately 10 kHz and one at 20 kHz in good agreement with the above simulation. The Q-factor of the 20 kHz resonance is approximately 8. Figure \ref{fig2}c) shows the measured resonance frequency of the QTF, when the top of the casing has been removed, thus at atmospherically pressures and at a temperature of 25 $^\circ$C. The measured resonance frequency and Q-factor at atmospheric pressure were 19.990 kHz and approximately 4500, respectively.

\section{Experimental Setup}

\begin{figure}[htbp]
\centering
\fbox{\includegraphics[width=\linewidth]{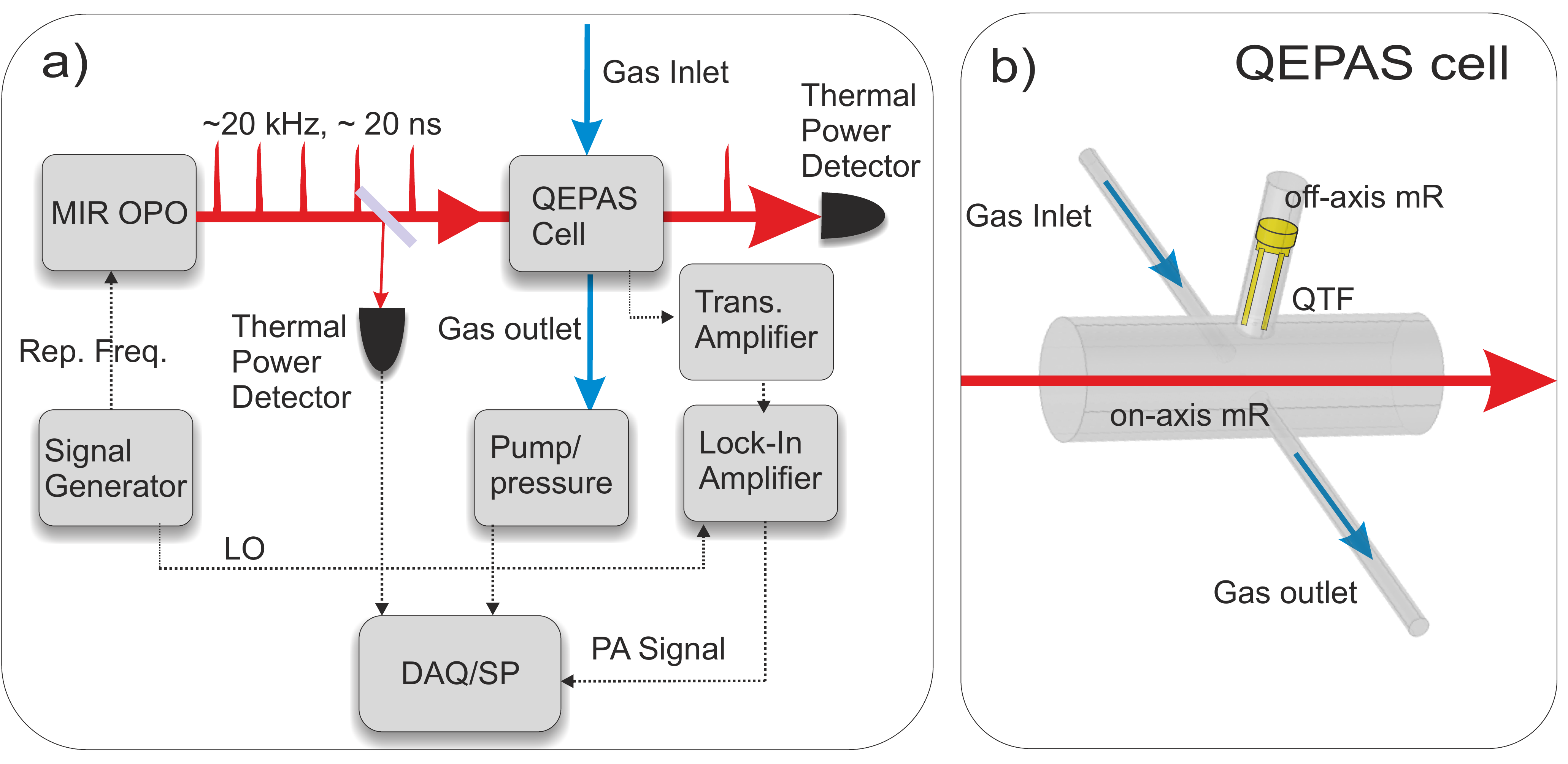}}
\caption{a) Block diagram of the important parts of the experimental setup. Lenses and mirrors are not shown. The experimental setup includes a nanosecond MIR OPO tunable between 3.1 and 3.8 $\mu$m, a novel mR system, a 20 kHz QTF, thermal detectors for monitoring the optical power, a signal generator for controlling the OPO repetition rate, pressure and temperature controllers for the QEPAS sensor and a lock-in amplifier connected to a 12 bit DAQ card. LO: Local oscillator, SP: Signal Processing. b) Layout of the QEPAS cell with the two acoustic coupled mRs.}
\label{fig1}
\end{figure}

The mR cell was constructed based on the simulations in high-density PTFE as depicted in Figure \ref{fig2}a). The PTFE material is used for several reasons. The PTFE walls decouples the in-phase background absorption signal from the PA gas signal due to thermal diffusion effects \cite{Lassen2014}. The MIR light beam enters and exits the on-axis mR via uncoated 3 mm thick calcium fluoride windows. For optimal transmission of the MIR light beam several lenses and mirrors are used (not shown in Figure \ref{fig1}a)). The measured optical transmission through the cell is 90$\%$. The QTF is placed inside the off-axis mR tube. The pressure is kept at atmospheric pressure at all times and the temperature of the cell is kept at 25$^\circ$C. This temperature has been chosen to optimize the acoustic coupling of PA signal generated in the on-axis tube to the off-axis QTF. The current signal from the QTF is first amplified by a transimpedance amplifier and then amplified by a pre-amplifier with a 1 kHz bandpass filter at 20 kHz before being processed with a lock-in amplifier and finally digitized with a 12 bit DAQ card.

For highly sensitive and selective trace-gas sensing it is desirable to have high energy sources with large wavelength tunability in the mid-infrared (MIR) region, where most molecules have strong vibrational transitions (fingerprint region) \cite{Sigrist2008}. A number of different light sources (QCLs, LEDs, DFBs, OPOs and more) have been reported used in QEPAS experiments \cite{Kohring2015,Spagnolo2015,Ma2013,Yi2014,Patimisco2014}. OPOs seem to be the optimal choice for providing large wavelength tunability, high energy, molecular selectivity and cost-effective device for the generation of infrared light in the 1.5 to 5 $\mu$m spectral range \cite{Petrov2015}. Therefore, a pulsed single mode mid-infrared (MIR) OPO has been developed. The pulse repetition rate is matched to the resonance frequency of the acoustic resonance of the QTF and the QEPAS senor. The light source is an OPO based on a 50 mm long PPLN nonlinear crystal with a fanned-out structure. The PPLN is placed inside a single-resonant cavity. The OPO is pumped at 1064 nm with diode-pumped nanosecond laser with an average output power reaching up to 27W. The Q-switch repetition rate can be changed continuously from 10 kHz to 80 kHz and the pulse duration from 7 to 50 ns. The maximum wavelength ranges for the signal and idler are 1.4 to 1.7 $\mu$m and 2.8 to 4.8 $\mu$m, respectively, and with output power ranging up to 4 W. The tunability of the OPO is achieved by the vertical translation of the fanned-out PPLN crystal, while the temperature of the crystal is kept at 30$^\circ$C. The system is kept single-mode using a 50$\mu$m thick etalon plate. This provides a bandwidth of around 0.1-1 cm$^{-1}$. A spectrometer is integrated in the system and the position of the stepping motors of the crystal mount, the etalon plate and the grating are controlled by a computer. In the present work the OPO is optimized for operation in the spectral region between 3.1 $\mu$m and 3.8$\mu$m and with an average output power of approximately 400 mW and pulse durations of $\sim$ 18 nanoseconds at a repetition rate of 19.99 kHz, thus matching the QTF and mR tubes.

\section{Methane measurements}

\begin{figure}[htbp]
\centering
\fbox{\includegraphics[width=\linewidth]{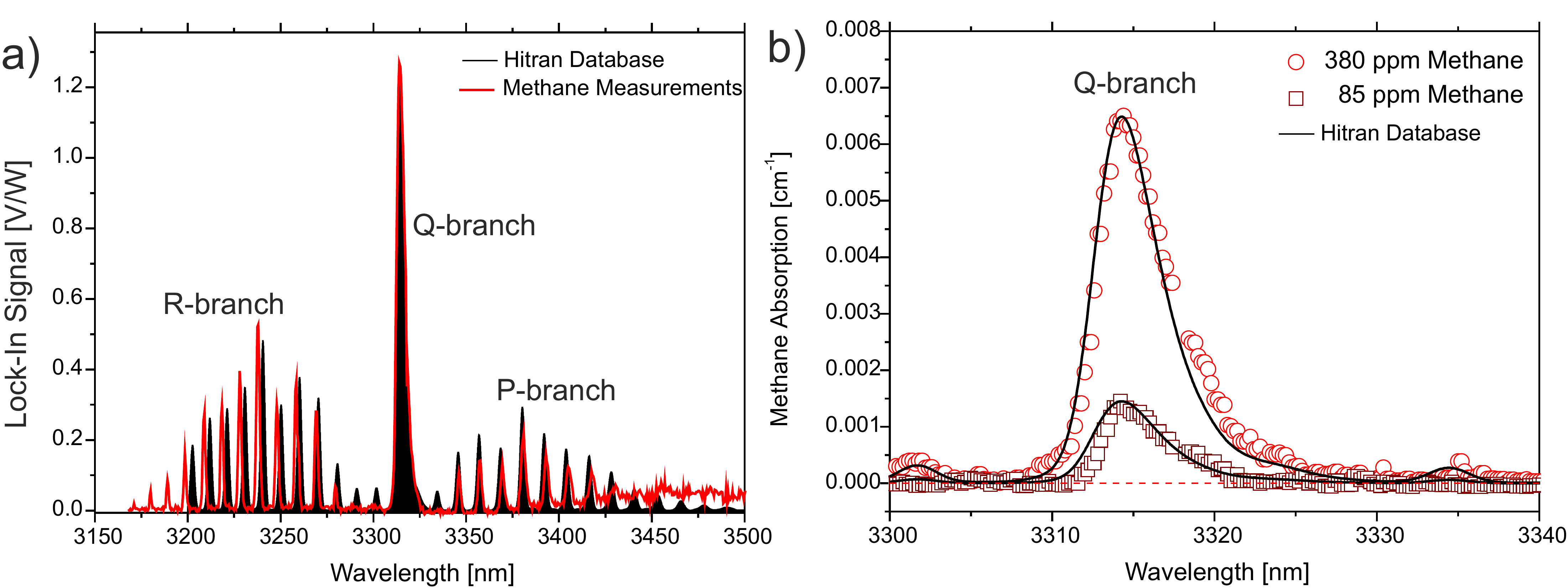}}
\caption{a) The red curve shows the measured methane spectrum with 100 ppm concentration and at 1 atm pressure. The black curve shows data from the Hitran database. b) Absorption per cm$^{-1}$ for two concentrations of methane. The points are experimental data and the solid lines are data from the Hitran database assuming a Gaussian instrument function with bandwidth of 1 cm$^{-1}$.}
\label{fig3}
\end{figure}

The experiments were performed by excitation of molecular ro-vibrational transitions of methane in synthetic dry air in the 3.1-3.5 $\mu$m wavelength region. Methane detection is of considerable interest since it is a major greenhouse gas and also considered as a potential biomarker for different gut and stomach inflammatory diseases and colorectal cancer \cite{Lourenco2014}. Figure \ref{fig3}a) shows the measured spectrum of methane (CH$_4$) with a 100 ppm concentration in synthectic air. We clearly see the R-, Q- and P-branch of methane. The Q-branch has a peak value of 1.28V. The data shown in Figure \ref{fig3}a) and b) were processed with a lock-in amplifier with a time constant of 300 ms, this corresponds to the energy accumulation time of the QTF at atmospheric pressure. The wavelength of the OPO was changed in steps of 0.5 nm. The spectrum has been normalized with respect to the optical power, which changes approximately 20$\%$ over the whole scanning range. The methane spectrum is compared with a spectrum from the Hitran database assuming same experimental conditions and with a Gaussian instrument function with bandwidth of 1 cm$^{-1}$. In order to estimate the resolution bandwidth of the sensor we only scan the Q-branch, the data is shown in Figure \ref{fig3}b).  From the data shown in Figure \ref{fig3}a) and b) we find a very good agreement with Hitran database, thus we estimate our complete QEPAS sensor resolution bandwidth to be approximately 1 cm$^{-1}$. Figure \ref{fig3}a) shows that over the whole scanning range their is a small offset between the measured spectra and the Hitran data. This is due to hysteresis in the stepping motors of the OPO system that gives rise to a small wavelength misalignment of the spectrometer, however, this mismatch of the wavelength axis can be compensated for in the data processing. To our knowledge these spectroscopic measurements are the first to combine QEPAS with a high power widely tunable nanosecond pulsed MIR OPO. We therefore believe that the QEPAS sensor will be very useful for environmental, industrial, and biological monitoring, where multiple gaseous pollutants and aerosols need to be monitored simultaneously.

\begin{figure}[htbp]
\centering
\fbox{\includegraphics[width=\linewidth]{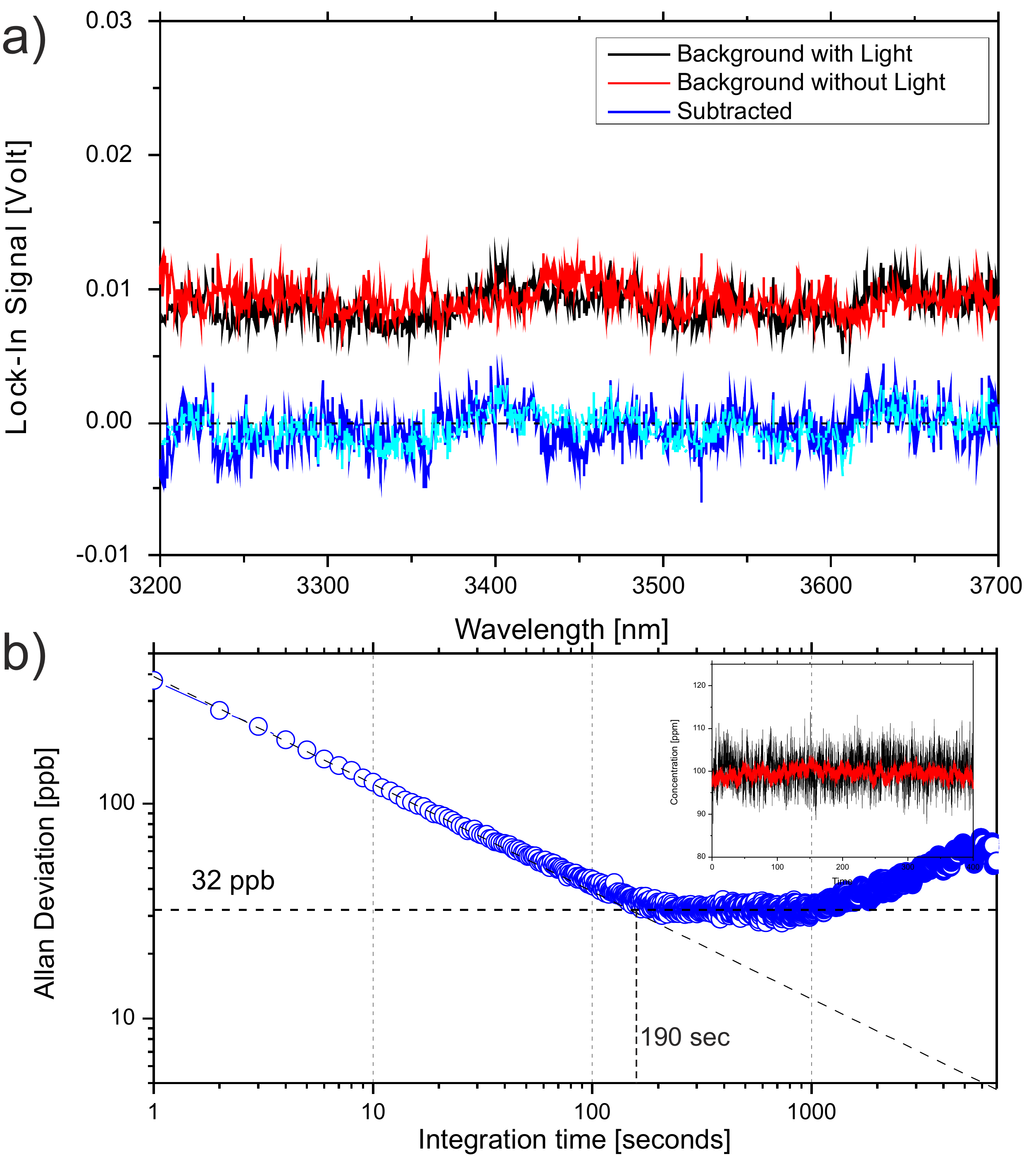}}
\caption{a) The background noise with and without light. The noise level is 0.0098V. The blue curves are the subtracted curves. b) Allan deviation analysis of the lock-in signal with 10 seconds integration and measured at 3.314 $\mu$m. The inset figure is the measured time trace for 300 ms and 10s integration. Detection sensitivity at optimum integration time is 32 ppbv@190s.}
\label{fig4}
\end{figure}

In order to have a practical and versatile QEPAS sensor for trace sensing, having a small resolution bandwidth and high tunability is not enough high sensitivity is also needed. In the following we estimate the sensitivity by performing measurements on the peak of the Q-branch. The main limiting factor for the detection sensitivity for PAS sensors is background noise signals. The background noise in QEPAS systems are due to electromagnetic pickup, stray light from the laser, cell absorption, flow, and QTF thermal noise. In order to achieve high detection sensitivity all these noise sources need to be dealt with. Different methods have so far been reported for effective background signal suppression of QEPAS sensors, for example the modulation cancelation method, which utilizes two modulated laser sources using an off-axis QTF as demonstrated in ref.\cite{Zheng2016_OE}. However, this method is slightly more complicated than the simple approach demonstrated in here. The background noise was measured by flushing the cell with pure air followed by a wavelength scan. Figure \ref{fig4}a) shows the measured background noise signal. We find that the background noise, with and without light, has the same magnitude and standard deviation. We therefore conclude that the background noise is not affected by stray light or cell absorption noise. This is shown in Figure \ref{fig4}a) where the background noise traces with and without light are subtracted. We find that the mean value of the subtracted trace is approximately 0. This means that the background noise arises from potential electromagnetic pickup noise, amplifier noise and probably mostly due to thermal noise of the QTF. The detection sensitivity can therefore simply be estimated by taking the ratio between the maximum methane signal and the QTF thermal noise level. The maximum methane signal can be found in Figure \ref{fig3}a) and b), where the Q-branch (at 3.314 $\mu$m) for methane with a 100$\pm 5$ ppm concentration has a value of 1.28 V and the noise level in Figure \ref{fig4}a) is 0.0098 V. The signal to noise ratio (SNR) is approximately 130, thus the noise equivalent detection sensitivity (NEDS) is 0.8 $\pm 0.04$ ppm. The obtained NEDS is approximately one order of magnitude lower than state-of-art methane measurements conduct in the 3.3 $\mu$m wavelength region \cite{Patimisco2014}. This is attributed to the QTF which has a slightly higher equivalent series resistance and lower Q-factor than the standard 32.7 kHz QTFs and the fact that the experiments were conducted in dry synthetic air which will lead to a strongly reduced PA signal \cite{Barreiro2011}. However note that by applying 4W of average MIR power and better design (higher Q-factor) of the mR tubes we believe that the NEDS can easily be improved to 20-40 ppb for methane making the sensor as sensitive as state-of-art QEPAS sensors \cite{Patimisco2014}.

The NEDS of 0.8 ppm describes the QEPAS sensor performance on a short time scale, however characterization of long-term drifts and signal averaging limits is very important for sensors. The optimum integration time and detection sensitivity are therefore determined using an Allan deviation analysis. The OPO was locked to the Q-branch absorption line peak while the methane concentration is maintained constant. The data consist of 100k points and was recorded over 12 minutes. Figure \ref{fig4}b) shows the Allan deviation analysis for the QEPAS data processed with a lock-in amplifier with a integration time constant of 10 seconds. The figure inset in Figure \ref{fig4}b) represents the time traces. The analysis shows that the detection sensitivity at optimum integration time is 32 ppbv@190s for methane measured at 3.314 $\mu$m. This means that white noise remains the dominant noise source for 190 s. This is where the ultimate detection limit is reached and after 1000 s the instrumental drift starts dominating.

\section*{Conclusion}
In conclusion, we have demonstrated a novel QEPAS configuration using acoustic coupled micro-resonators with an off-axis 20 kHz quartz tuning fork (QTF). The QEPAS system is pumped resonantly by a nanosecond pulsed single-mode MIR OPO. Different spectral features of methane is resolved and the R-, Q- and P-branchs are clearly identified. From the comparison of the methane spectra with the Hitran spectra we conclude that the QEPAS instrument resolution bandwidth is approximately 1 cm$^{-1}$. The SNR was 130 for these spectropic measurements, which lead to a detection sensitivity of approximately 0.8 ppm. However by applying optimum integration time the sensitivity can be increased to 32 ppbv@190s. We believe that the tunability and sensitivity demonstrated here is sufficient to make the QEPAS sensor system very useful for environmental, industrial, and biological monitoring, where multiple gaseous pollutants and aerosols need to be monitored simultaneously. Further improvement of the QEPAS sensor will be made to make the complete system more compact and robust in order perform in situ monitoring outside the safe environment of the laboratory.

\section*{Funding Information}
We acknowledge the financial support from EUREKA (Eurostars program: E9117–NxPAS) and the Danish Agency for Science Technology and Innovation. We would like to thank Anders Brusch for fruitful discussions.

\end{document}